\documentclass{mn2e}
\usepackage{epsf}
\usepackage{graphicx}

\def\etal{{et~al.}}

\title[XTE J1118+480 in quiescence]
{Doppler and modulation tomography of XTE J1118+480 in quiescence}

\author[D.E. Calvelo \etal]
{D.E. Calvelo,$^{1,2}$ S.D. Vrtilek,$^{1}$ D. Steeghs,$^{1,3}$ M.A.P. Torres,$^{1}$
J. Neilsen,
$^{1,4}$
\newauthor 
A.V. Filippenko,$^{5}$ and 
J.I. Gonz\'{a}lez Hern\'{a}ndez,$^{6,7}$\\ 
$^{1}$ Harvard-Smithsonian Center for Astrophysics, 60 Garden
Street, Cambridge, MA 02138, USA\\ 
$^{2}$ School of Physics and Astronomy, University of Southampton,
Highfield, Southampton, SO17 1BJ, UK\\
$^{3}$ Department of Physics, University of Warwick, Coventry, CV4
7AL, UK\\ 
$^{4}$ Harvard University,
Department of Astronomy, 60 Garden Street, MS-10, Cambridge, 
MA 02138, USA\\
$^{5}$ Department of Astronomy,
University of California, Berkeley, CA 94720-3411, USA\\
$^{6}$ Observatoire de Paris-Meudon, GEPI, 5 place Jules
Jannsen, 92195 Meudon Cedex, France\\
$^{7}$ CIFIST Marie Curie Excellence Team}

\date{\today}

\volume{000}

\pagerange{\pageref{firstpage}--\pageref{lastpage}} \pubyear{2006}

\begin{document}

\label{firstpage}

\maketitle
\begin{abstract}
We present Doppler and modulation tomography of the X-ray nova XTE J1118+480
with data obtained during quiescence using the 10-m Keck II telescope.
The hot spot where the gas stream hits the accretion disc is seen in
H$\alpha$, H$\beta$, He {\sc i} $\lambda$5876, and Ca {\sc ii} 
$\lambda$8662, thus verifying the presence of continued mass transfer 
within the system. 
The disc is clearly
seen in H$\alpha$ and Ca {\sc ii} $\lambda$8662.  
We image the
mass-donor star in narrow absorption lines of 
Na {\sc I} $\lambda\lambda$5890, 
5896, 8183, 8195 and Ca {\sc ii} $\lambda$8662, implying an 
origin from the secondary itself rather than the interstellar medium. 
We also detect deviations in the
centroid of the double peak of H$\alpha$ akin to those found by 
Zurita~\etal~2002 
suggesting disc eccentricity.

\end{abstract}


\section{Introduction}

XTE J1118+480 is a member of a subclass of low-mass X-ray binaries 
(LMXBs) known
as X-ray novae (XRNe) or soft X-ray transients. These are systems 
in which
the low-mass companion fills its Roche lobe, resulting in
accretion onto the heavier primary (a neutron star or black hole). The
XRNe in particular undergo significant luminosity variations over
time; they occur as ``outburst'' events involving rapid
increases in X-ray light by many orders of magnitude over timescales of
days, coupled with a noticeable increase in optical light. Systems can
remain in outburst for months before declining into an epoch of
quiescence where they may remain for decades before undergoing another
similar event (e.g., Chen, Shrader, \& Livio 1997).

The cause and processes behind such events are yet to be made
clear. One of the more favored explanations involves the application
of disc instability models (Meyer \& Meyer-Hofmeister 1981). Such theories were
originally conceived for discs in cataclysmic variable systems but 
can also be applied 
to discs that form around the primaries of LMXBs. The quiescent
state can be regarded as a period where the disc gains mass and
evolves until it reaches a critical point and destabilizes, resulting
in large quantities of mass being dumped onto the primary in a short
period of time, in turn causing the outburst we see.

XTE J1118+480 was discovered during an outburst (Remillard~\etal~2000) 
on
2000 March 29 (UT dates are used throughout this paper) by the Rossi 
X-ray Timing Explorer (RXTE; Jahoda~\etal~1996) All-Sky
Monitor (ASM; Levine~\etal~1996). Its optical counterpart 
(brightening to $V \approx 12.9$ mag
during outburst) was discovered in the same 
year by Uemura, Kato, \& Yamaoka  (2000a), corresponding to an 18.8
mag star in USNO catalogues. A 4.1~hr period was detected by several
independent investigations (Patterson~\etal~2000; Uemura~\etal~2000b), and 
distortions within light curves were attributed to
superhumps (Uemura~\etal~2000c; Zurita~\etal~2002, henceforth Zu02;
Zurita~\etal~2006; Torres~\etal~2004, henceforth To04). Such phenomena
theoretically occur when the accretion disc expands to encompass tidal
resonance radii, triggering precessional effects on the
disc due to the gravitational pull of the secondary star (Whitehurst \&
King 1991). The deformed disc then causes intensity variations as its projected area varies with phase.

XTE J1118+480 was also observed in a ``mini-outburst'' state
in 2005 by Zurita~\etal~(2006), who noted that the system exhibited
superhump activity during the mini-outburst before settling back into
quiescence.

Subsequent observations since the discovery of XTE J1118+480 have yielded 
well-constrained system parameters. Radial-velocity measurements of the
secondary star carried out during the post-outburst period gave
a mass function of $\sim$ 6~M$_{\odot}$ (McClintock~\etal~2001a; 
Wagner~\etal~2001; Gonz\'{a}lez Hern\'{a}ndez~\etal~2006, hereafter 
GH06), providing 
evidence that the compact object is a black
hole. Since superhump modulations can alter the apparent ephemeris,
observations of XTE J1118+480 in quiescence (which presumably 
lacks superhump 
activity)
have allowed for refinement of many system parameter measurements including the
period (Gelino~\etal~2006; Gonz\'{a}lez Hern\'{a}ndez~\etal~2008, 
hereafter 
GH08). Gelino~\etal~(2006) improved the determination of the inclination 
angle (to $68^\circ \pm 2^\circ$) which constrained the mass of the primary 
to $8.53 \pm 0.60$~M$_{\odot}$, confirming its identification 
as a black hole. System parameter values, including
those used in calculations herein, are listed in Table 1.

\begin{table}
\caption{XTE J1118+480 System Parameters} 
\begin{tabular}{|r|l|}
\hline
Orbital period & 0.16995 $\pm$ 0.00012 d $^{(1)}$ \\
HJD phase 0 & 2453049.93346 $\pm$ 0.00007 d $^{(1)}$ \\
Mass Ratio $q$ (M$_{2}$/M$_{1}$) & 0.024 $\pm$ 0.009 $^{(3)}$ \\
 & 0.0435 $\pm$ 0.0100 $^{(2)}$ \\
Inclination & $68^\circ \pm 2^\circ$ $^{(2)}$ \\
$v$\,sin\,$i$ & $96^{+3}_{-11}$ ${\rm km}\:{\rm s}^{-1}$  $^{(3)}$ \\
Primary Object Mass & 8.53 $\pm$ 0.60 M$_{\odot}$ $^{(2)}$ \\
Mass Function & 6.27 $\pm$ 0.04 M$_{\odot}$ $^{(1)}$ \\
$K_2$ & 708.8 $\pm$ 1.4 km s$^{-1}$ $^{(1)}$ \\
Systemic velocity $\gamma$ & 2.7 $\pm$ 1.1 km s$^{-1}$ $^{(1)}$ \\
Distance & 1.72 $\pm$ 0.10 kpc $^{(2)}$ \\
Secondary spectral type & K5 V $^{(1)}$ \\
\hline
\end{tabular}
Note: $^{(1)}$Gonz\'{a}lez Hern\'{a}ndez~\etal~(2008). 
$^{(2)}$Gelino~\etal~(2006). $^{(3)}$See Section 2. The Gelino~\etal~ 
parameters were calculated in conjunction with a different mass ratio 
and thus we include both values for reference.
\end{table}

Doppler tomography (Marsh \& Horne 1988) 
has been applied to XTE J1118+480 several times
since its discovery. Dubus~\etal~(2001) produced He {\sc II} and H$\alpha$
tomograms from observations taken on 2000 April 20, while the
system was in outburst. A bright region in the maps was attributed
to a hot spot where the accretion stream interacts with the
disc. Torres~\etal~(2002, henceforth To02) produced tomograms for
the nights of
2000 March 31, April 12 and 29, and May 25 (during the same outburst),
revealing a disc as well as a ``bow-shaped'' emission region whose
position suggested ties with the accretion stream or a stream-disc
interaction.

The most recent tomograms produced prior to those in this paper are found
in To04 and Elebert~\etal~(2006). The To04 observations were
taken on the nights of 2000 Dec. 29, 2001 Jan.
26, and 2003 Jan. 2 and 3 in order to cover decline into quiescence and
quiescence itself. A possible hot spot was observed in both the 2000
and 2001 images, as well as evidence for a precessing eccentric disc. Also
apparent was a second bright region perhaps due to accretion-stream
overflow. Most
interestingly, although a disc is apparent, no hot-spot region
was detected in the To04 2003 tomogram. The asymmetric component of
the 2003 map revealed
H$\alpha$ emission centered on the companion's 
Roche lobe rather than at a
position which could be attributed to a hot spot. The implication is
that a change in accretion rate had occurred since the earlier
observations. Surprisingly, Doppler tomograms by Elebert~\etal~(2006) 
of observations taken during the 2005 mini-outburst also 
do not show a significant hot spot in H$\alpha$.
The timings of relevant XTE J1118+480 optical observations are shown
in Figure 1 against the system's X-ray intensity.

\begin{figure}
\centerline{\includegraphics[width=3.2in]{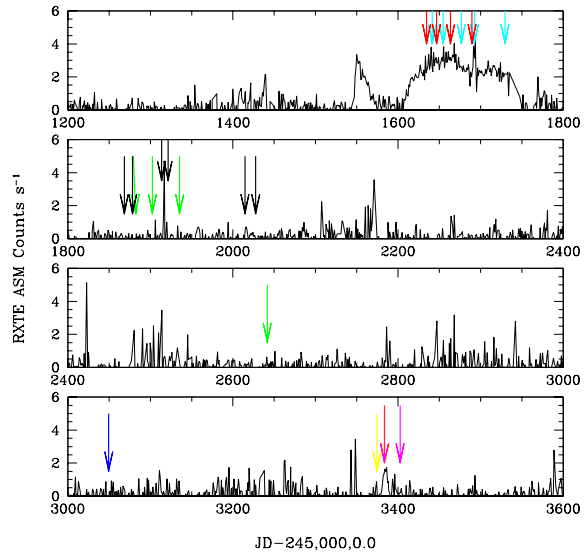}}
\caption[RXTE ASM light-curve data for XTE J1118+480]{RXTE ASM light-curve
data for XTE J1118+480. Data were retrieved using the ASM data website at
http://xte.mit.edu/ASM\_lc.html. Arrows indicate the times of optical 
observations. 
Cyan = Dubus~\etal~(2001); black = Zurita~\etal~(2002; no tomography);
red = Torres~\etal~(2002); green = Torres~\etal~(2004); blue = current
work; yellow = Zurita~\etal~(2006; no tomography); magenta = 
Elebert~\etal~(2006).}
\end{figure}

In this paper we produce Doppler and modulation  
tomography maps of
XTE J1118+480 from Keck II observations taken on 2004 Feb. 14, when the
system was well into quiescence. One of the shortcomings of Doppler
tomography is that it assumes emission is constant over the
orbit. Modulation tomography (Steeghs 2003) addresses this 
problem and hence can
reveal the structure of accretion flow in greater detail. Our
observations represent greater than a factor of 4 improvement in spectral
resolution over previous observations used for 
tomography of XTE J1118+480. 
The higher spectral resolution, in addition to improved orbital phase resolution, enabled us to produce detailed modulation
tomography maps of H$\alpha$, the strongest line. We
search for evolutionary trends by comparing our tomograms with those
taken at different outburst phases.

\section{Observations}

The observations used in this paper were initially described by GH06.
They consist of 74 medium-resolution ($\lambda / \Delta
\lambda$ $\approx$ 6,000) spectra of XTE J1118+480 taken on 2004 Feb. 14,
using the Echellette Spectrograph and Imager (Sheinis et
al. 2002) on the 10-m Keck II telescope. Table 2 lists the observations and the
associated orbital phase calculated using the ephemeris determined
by GH08 and listed in Table 1. All spectra were reduced in a
standard manner as described by GH06.

\begin{table}
\caption{Journal of Observations on 2004 February 14.}
\begin{footnotesize}
\begin{tabular}{l l l | | l l l}
\hline
Record & UTC  & Phase & Record & UTC  & Phase \\
\hline
1  &   7.86802   &   -0.5956 & 38  &  11.80850  &   0.3705 \\      
2  &   7.97901   &   -0.5684 & 39  &  11.91085  &   0.3956 \\      
3  &   8.08408   &   -0.5426 & 40  &  12.01332  &   0.4207 \\      
4  &   8.19076   &   -0.5165 & 41  &  12.11583  &   0.4458 \\      
5  &   8.30549   &   -0.4883 & 42  &  12.21775  &   0.4708 \\      
6  &   8.41041   &   -0.4626 & 43  &  12.35703  &   0.5050 \\      
7  &   8.51713   &   -0.4365 & 44  &  12.45861  &   0.5299 \\      
8  &   8.65651   &   -0.4023 & 45  &  12.56290  &   0.5555 \\      
9  &   8.76902   &   -0.3747 & 46  &  12.66709  &   0.5810 \\      
10  &   8.87278  &   -0.3493 & 47  &  12.76956  &   0.6061 \\      
11  &   8.97658  &   -0.3238 & 48  &  12.87331  &   0.6316 \\      
12  &   9.08021  &   -0.2984 & 49  &  12.97488  &   0.6565 \\      
13  &   9.17981  &   -0.2740 & 50  &  13.07572  &   0.6812 \\      
14  &   9.27935  &   -0.2496 & 51  &  13.17689  &   0.7060 \\      
15  &   9.38655  &   -0.2233 & 52  &  13.30560  &   0.7375 \\      
16  &   9.51406  &   -0.1920 & 53  &  13.41519  &   0.7644 \\      
17  &   9.61566  &   -0.1671 & 54  &  13.51554  &   0.7890 \\      
18  &   9.71547  &   -0.1426 & 55  &  13.61731  &   0.8140 \\      
19  &   9.81456  &   -0.1184 & 56  &  13.71921  &   0.8389 \\      
20  &   9.92418  &   -0.0915 & 57  &  13.82052  &   0.8638 \\      
21  &  10.02500  &   -0.0668 & 58  &  13.92143  &   0.8885 \\      
22  &  10.12409  &   -0.0425 & 59  &  14.02387  &   0.9136 \\      
23  &  10.22280  &   -0.0183 & 60  &  14.13096  &   0.9399 \\      
24  &  10.36291  &   0.0161 & 61  &  14.27274  &   0.9746 \\      
25  &  10.46241  &   0.0405 & 62  &  14.39129  &   1.0037 \\      
26  &  10.56251  &   0.0650 & 63  &  14.49344  &   1.0288 \\      
27  &  10.66197  &   0.0894 & 64  &  14.59468  &   1.0536 \\      
28  &  10.76201  &   0.1139 & 65  &  14.71055  &   1.0820 \\      
29  &  10.86116  &   0.1382 & 66  &  14.80933  &   1.1062 \\      
30  &  10.96110  &   0.1627 & 67  &  14.91224  &   1.1314 \\      
31  &  11.06149  &   0.1874 & 68  &  15.18856  &   1.1992 \\      
32  &  11.16561  &   0.2129 & 69  &  15.28683  &   1.2233 \\      
33  &  11.26717  &   0.2378 & 70  &  15.38589  &   1.2476 \\      
34  &  11.39890  &   0.2701 & 71  &  15.48483  &   1.2718 \\      
35  &  11.50254  &   0.2955 & 72  &  15.58512  &   1.2964 \\      
36  &  11.60421  &   0.3204 & 73  &  15.68769  &   1.3216 \\      
37  &  11.70602  &   0.3454 & 74  &  15.78674  &   1.3458 \\   
\hline
\end{tabular}
Note: The orbital phase values in columns
  3 and 6 are computed using the orbital solution from GH08.

\end{footnotesize}
\end{table}

\section{Spectral analysis}

\subsection{Spectra}

The smearing of lines due to the motion of the star during a given
exposure was incorrectly
determined by Gonz\'{a}lez Hern\'{a}ndez~\etal~(GH06).
However, this has a small impact in the derived rotational
velocity. 
We have determined the smearing to be in the range 2--90 km
s$^{-1}$. 
Our new value for the rotational velocity is $v$\,sin\,$i=96^{+3}_{-11}$ 
${\rm km}\:{\rm
  s}^{-1}$ (originally $=100^{+3}_{-11}$ ${\rm km}\:{\rm
  s}^{-1}$ in GH08), where the uncertainties have been derived by 
assuming extreme
cases for the linearized limb-darkening $\epsilon = 0-1$. This
rotational velocity combined with the GH08 value of the secondary-star 
radial-velocity 
semi-amplitude, $K_2$, implies a binary mass ratio
$q=0.024\pm0.009$ (originally $=0.027\pm0.009$).

The reduced spectra were examined for spectral features suitable
for use in tomography studies.  The strongest lines
(H$\alpha$, H$\beta$, and Ca {\sc ii}) show significant structure 
in their profiles,
and are thus good candidates.  The sensitivity of our data also 
allowed us to study the much weaker lines of He {\sc i} $\lambda$5876 and
Na {\sc I} $\lambda\lambda$5890, 5896, 8183, 8194.

The spectra were analyzed using the MOLLY software developed
by Tom Marsh. We
concentrate on six spectral regions that cover the brightest lines
(H$\alpha$, H$\beta$, He {\sc i}/Na {\sc I}, Na {\sc I} absorption near 
$\lambda$8200, and several lines of Ca {\sc ii}). 
Sample spectra from
two orbital phases are shown in Figure 2: the double-peak
structure associated
with rotating disc emission is clearly evident in H$\alpha$.
As the strongest visible
emission line (average equivalent width 
of about 0.5~\AA), H$\alpha$ is the best defined.
The location of 
H$\beta$ near the blue end of the spectrum 
(where device sensitivity begins
to drop) and its relatively low intensity (factor of $\sim$4
weaker than H$\alpha$; average equivalent width around 0.12~\AA) means 
that there is substantially more noise and 
the double-peak structure is harder to detect.
The double-peak structure is not resolved in He {\sc i}, and is blended
in Ca {\sc ii} because of overlap between closely spaced emission and
absorption features.

\begin{figure}
  \centerline{\includegraphics[width=3.2in]{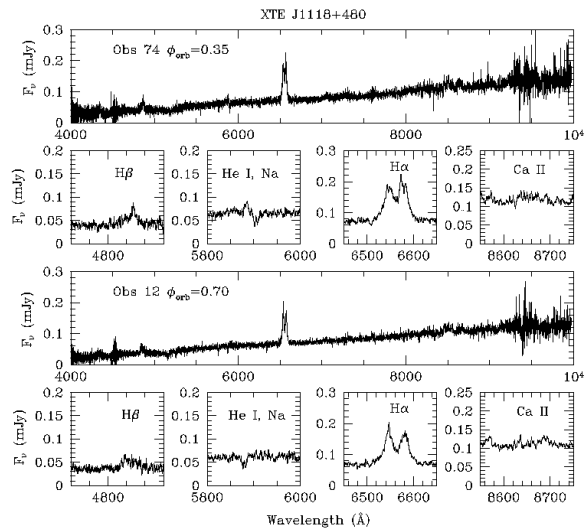}}
\caption[Sample spectra.]{Sample spectra from two orbital phases with 
close-ups of some
of the line features.  Upper panels: orbital phase 0.35.  Lower panels:
orbital phase 0.70.  Data have been smoothed with a boxcar of width 7 pixels.}
\end{figure}

\subsection{H$\alpha$ Variability}

The emission lines of XTE J1118+480 are known to exhibit strong
asymmetric structure as well as variations of emission-line centroid
position over time scales far longer than the orbital period (To02;
Zu02). Unlike absorption lines which remain centered at the same
wavelength (with only the small deviation caused by $\gamma$ = 2.7 km
s$^{-1}$), the repeating variance of these emission-line shifts means
they cannot be attributed to the systemic velocity.  Zu02 fitted a
three-component Gaussian (consisting of a broad base and two narrow
peaks) to their H$\alpha$ profiles, averaged over each night.
They found that the positions of the blue and red peaks are not
equidistant from the expected rest wavelength and measured offsets up
to 300 km~s$^{-1}$.  Fitting a sinusoid to the H$\alpha$ offsets they
measured on six nights, they found the variations to be consistent
with either the precession period of $\sim$52~days that they inferred
from measurement of a superhump period or half the precession period
($\sim$ 26 days; see their Fig. 5).  H$\alpha$ offsets measured by
To04 (see their Fig. 5) two years after the Zu02 observations were
consistent with the precession period but ruled out the 26 day
period. Simulations by Faulkes~\etal~ (2004) give convincing evidence
that such long time-scale modulations of broad emission lines can be
directly linked to the precession of an eccentric, asymmetric
disc. In essence, the precession alters the angle at which we observe
the elliptical orbit of matter in the disc, thereby causing the
velocities we observe at the blue and red ``edges'' of the disc to
oscillate with the precession period.

\begin{table}
\caption{H$\alpha$ centroid offsets.} 
\begin{center}
\begin{tabular}{lcccc}
\hline
Date & $V_{\rm b}$& $V_{\rm r}$& $V_{\rm r} - V_{\rm b}$ & $V_{\rm off}$ \\
\hline
2000 Mar 31 &-643$\pm$24&598$\pm$20&1241$\pm$31&-22.5$\pm$15.5\\
2000 Apr 12 &-601$\pm$29&521$\pm$17&1122$\pm$34&-40.0$\pm$17\\
2000 Apr 29 &-576$\pm$14&620$\pm$25&1196$\pm$29&22$\pm$14.5\\
2000 Nov 20 &-750$\pm$50&950$\pm$50&1700$\pm$70&100$\pm$35   \\ 
2000 Nov 30 &-1050$\pm$50&600$\pm$50&1650$\pm$70&-225$\pm$35   \\ 
2000 Dec  4 &-1163$\pm$34&664$\pm$14&1827$\pm$37&-250$\pm$18.5\\
2000 Dec 24 &-644$\pm$16&1021$\pm$23&1665$\pm$28&188$\pm$14\\
2000 Dec 29 &-651$\pm$8&1120$\pm$17&1771$\pm$19&234$\pm$9.5\\
2001 Jan  4 &-600$\pm$50&1100$\pm$50&$1700\pm$70&250$\pm$35   \\ 
2001 Jan 12 &-900$\pm$50&650$\pm$50&1550$\pm$70&-125$\pm$35   \\ 
2001 Jan 26 &-1134$\pm$12&627$\pm$7&1761$\pm$14&-254$\pm$7\\
2001 Apr 15 &-700$\pm$50&1000$\pm$100&1700$\pm$110&150$\pm$55   \\ 
2001 Apr 16 &-500$\pm$50&950$\pm$50&$1450\pm$70&225$\pm$35   \\ 
2001 Apr 27 &-800$\pm$50&800$\pm$50&1600$\pm$70&0$\pm$35   \\ 
2001 Apr 28 &-850$\pm$50&750$\pm$50&1600$\pm$70&-50$\pm$35   \\ 
2003 Jan  2 &-952$\pm$6&811$\pm$8&1763$\pm$10&-71$\pm$5\\
2003 Jan  3 &-1009$\pm$6&823$\pm$8&1832$\pm$10&-93$\pm$5\\
2004 Feb 14 &-786$\pm$4&894$\pm$4&1680$\pm$6&55$\pm$3\\
2005 Jan 12 &-662$\pm$11&681$\pm$11&1343$\pm$16&9$\pm$8\\ 
2005 Feb  1 &-843$\pm$17&809$\pm$11&1652$\pm$20&-17$\pm$10\\
\hline
\end{tabular}
\end{center}
Note: Offsets are determined using spectra 
averaged over each night (for nights when there
is coverage over a full binary orbit).
All velocities are in km s$^{-1}$.  $V_{\rm off} = (V_{\rm r} + V_{\rm b})/2$. 
Numbers before 2004 are from To02, Zu02, and To04. The values with 
particularly large uncertainties are those read off of Figure 7 in Zu02. 
\end{table}
  
We measured the H$\alpha$ centroid offset of our observations, 
which were obtained about one year after the
To04 observations, in two ways.  First we averaged all 74 observations
and fitted three Gaussians to the average profile 
as was done by Zu02 and
To04 (after accounting for $\gamma$).  We also 
measured the central velocity and peak-to-peak separation for 
each spectrum 
individually and then  
fit to the system's radial-velocity to determine the offset.
Both methods agree to well within one sigma.  
We find an overall 
redward deviation from the rest 
wavelength (6562.87~\AA)
of 1.2~\AA\ that corresponds to an offset of $54.6 \pm 4$ km~s$^{-1}$.

We searched the literature for further
examples of H$\alpha$ centroid offset measurements.  To the
8 points originally plotted by Zu02 and the 4 points added by
To04, we add 3 points from To02 and an additional 2 points from To04
(not included in their Fig. 5 but appearing in their Table 3).
We also add 2 points from reanalyzing data obtained by Elebert~\etal~
(2006)
approximately one year after our Keck observations,
for a total of 20 measurements of the H$\alpha$ offset.
We list these values in Table 3 and plot them in Figure 3.
We use our offset value to refine the precession period to 
$P_{\rm prec} = 51.42 \pm 0.05$~days. 

\begin{figure}
  \centerline{\includegraphics[width=3.2in]{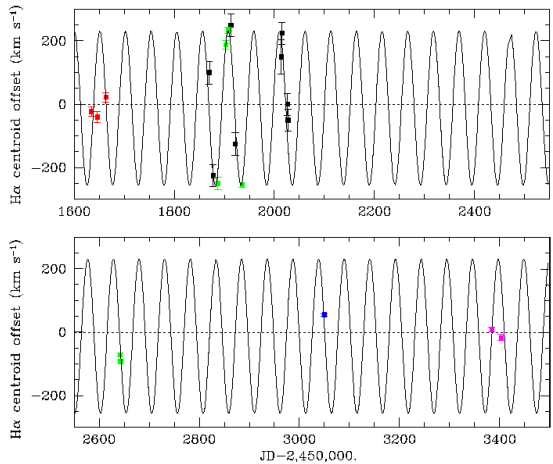}}
\caption[H$\alpha$ deviations.]{H$\alpha$ centroid offsets. The black points
represent Zu02, red points To02, and green points To04 measurements; 
current observations are in blue; and magenta points are calculated from 
data presented by Elebert~\etal~(2006).
A sinusoid having a period of 51.42~days (with amplitude and $T_{0}$ 
from To04) is overplotted.}
\end{figure}

We do not have sufficient points to perform a periodogram, but find by
trial and error that a period of 51.42 days (maintaining the amplitude and
$T_{0}$ as determined by To04) gives the least $\chi^{2}$ deviation
for all of the measured values:

\vspace{10pt}
$V_{\rm offset} = 243.53 \, {\rm sin} [(2\pi/P_{\rm prec})(t-T_0)] - 12.56, ~~~~~(1)$ 
\vspace{10pt}

\noindent
where $T_0$ = JD 2,451,870.01. Comparing this fit to the observation
times displayed in Figure 1, we would be tempted to believe that
precession remains stable over long periods of time regardless of the
system's state. Since the disc of J1118+480 is relatively small (with
larger peak-to-peak velocities) and has been shown to vary in size,
such unvarying precession behaviour (in both semi-amplitude and
period) would be surprising; all tidal models are very sensitive to
the size of the accretion disc.  However, the sparse and irregular
data spacing makes the fit of a single precession period,
semi-amplitude, and phase offset highly uncertain, especially if we
consider the possibility that these parameters vary with time. Any
existing variation can only be investigated via a greater sampling of
data points from a range of different epochs, specifically different
precession phases.

The peak-to-peak velocity separation ($V_{\rm r} - V_{\rm b}$) is 
noticeably smaller during outburst
(2000 Mar--Apr and 2005 Jan 12) than during quiescence. 
These peak-to-peak velocity changes are  
well known in cataclysmic variable stars: 
the disc is large (thus exhibiting a small peak-to-peak 
separation) during outburst and then 
shrinks (i.e., the peak separation widens) in quiescence.
We note that the Ca {\sc ii} $\lambda$8662 emission line also shows a 
strong disc signature in the form
of double-peak line profiles.  However, we were unable to 
obtain a satisfactory fit to 
the Ca {\sc ii} emission lines with the three-Gaussian model because of
blending with other lines.  We did not attempt any
fits for He {\sc i} and H$\beta$ since He {\sc i} also suffers interference
from nearby features and the disc signature in H$\beta$ is weak.

\begin{figure*}
\centerline{\includegraphics[width=7in,angle=0]{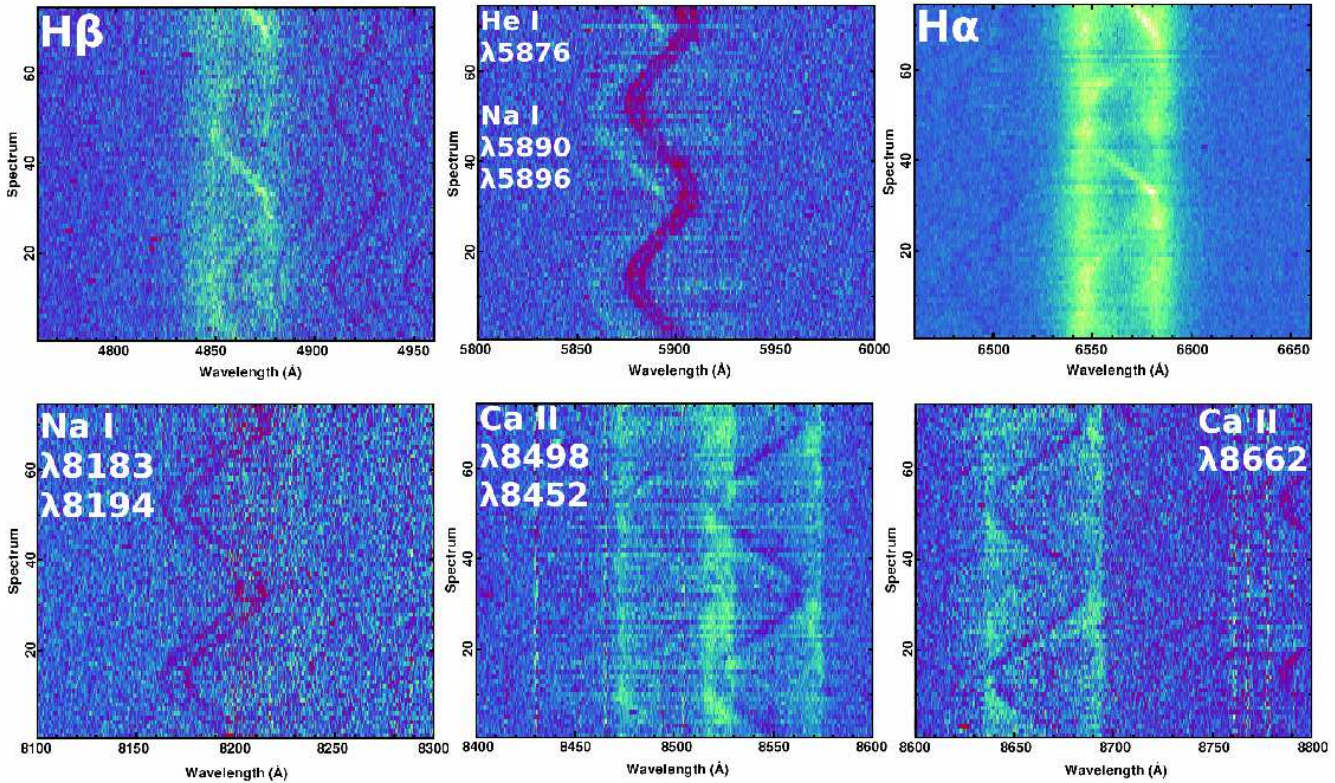}}
\caption[Trailed spectrograms.]{Trailed spectrograms. Upper left: The
  strong emission (green) feature is H$\beta$ and several weaker
  absorption (red) lines are visible to the right. Upper middle: He
  {\sc i} in emission and Na {\sc i} D lines in absorption; also an
  unidentified faint absorption feature to the left of He {\sc
    i}. Upper right: H$\alpha$ in emission and a faint absorption
  feature to its left. Lower left: Na {\sc i} $\lambda$8200 doublet in
  absorption. Lower middle: Ca {\sc ii} triplet seen in both emission
  and absorption.  Lower right: Another Ca {\sc ii} line in both emission
  and absorption. }
\end{figure*}

If the disc is large enough to encompass tidal resonance radii, then
we would have a plausible explanation for any effects implying the
existence of an
eccentric precessing disc. The peak-to-peak width of H$\alpha$ in the 2004 Feb 14 data is 
$36.93 \pm 0.1$~\AA, corresponding to a velocity difference of 
$1680 \pm 5.7$ km~s$^{-1}$. Following Dubus~\etal~(2001), we can 
make an estimate of the outer 
disc radius,
$$ R_{\rm d}/a = (1+q)(2K_2/\Delta V)^2, ~~~~~(2)$$

\noindent
where $\Delta V$ is the peak-to-peak separation. With the $\Delta V$
of H$\alpha$ and using $K_2$ and $q$ from Table 1, we get $R_{\rm d}/a
\approx 0.73 \pm 0.01$, close to the predicted tidal truncation radius
($\sim 0.9R_{\rm Roche}$, where $R_{\rm Roche}$ is the primary's
Roche-lobe radius) of $R_{\rm T}/a \approx 0.6$ for $q = 0.024$ (King
1995). It should be noted that the $s$-wave contribution to the
H$\alpha$ emission will have at least some effect on the measured
$\Delta V$ (as in H$\beta$) which would result in a small
overestimation of $R_{\rm d}$. The calculated outer radius exceeds
both the 3:1 tidal resonance radius $R_{3:1}/a \approx 0.48$, and the
2:1 resonance radius $R_{2:1}/a \approx 0.62$ (Whitehurst \& King
1991). Thus, even in quiescence, the disc appears to be large enough
to access its tidal resonance radii.

It should be stressed that Equation 2 is valid only when the orbital
motion of matter in the disc is Keplerian. Near outburst the disc is
expected to have grown to a point where it is susceptible to
destabilisation, and we have little understanding of the behaviour of
the disc matter. Hence, points in Table 3 that are measured during
outburst when used with Equation 2 give radii that are larger than the
binary separation. However, we are effectively left with two
scenarios; either the disc is small enough for us to believe the
assumptions of Keplerian motion and thus the use of Equation 2, or the
disc is sufficiently large (e.g., reaching the truncation radius) to
render Equation 2 useless. We only intend to show that the disc can access
resonance radii in order to explain any eccentricity effects in the
system, which for our calculations is true at both limits.

\section{Tomography}

Doppler imaging allows us to create a two-dimensional velocity-space
image of emission within a binary system using spectra and an
accurate ephemeris (Marsh \& Horne 1988). The spectra are 
considered projections 
of the disc in a single phase direction and are combined using a mathematical
inversion technique to produce a model that fits the data. We use the maximum
entropy method (MEM) to improve constraints on image
selection. MEM includes image entropy as a
variable to ensure that the resulting images are as smooth (and thus as 
physically realistic) as possible. We begin with an arbitrary starting 
image (e.g., uniform or a Gaussian peak) which is modified to fit the data
using a reduced $\chi ^{2}$ test; this process is iterated until a
satisfactory fit is achieved, while at the same time maximising the entropy of the reconstructed image.

The trailed spectrograms in Figure 4 (covering the wavelength ranges
of interest) reveal far more detail than the individual spectra. The
double-peak profiles of H$\beta$ become distinct and several weak
absorption features can be seen. More importantly, the weaker features
of He {\sc i} and Na {\sc I} are clearly resolved. The phase delay
seen between He {\sc i} and Na {\sc I} in the upper-middle panel of
Figure 4, as well as between H$\beta$ and the absorption features in
the upper-left panel of Figure 4, points to their originating from
different regions of the binary. The faint, barely detectable,
vertical lines at 5890~\AA\ and 5896~\AA, which show no Doppler effect
from the binary motion, appear to be in emission and thus cannot be
due to interstellar absorption but likely arise from residual
night-sky emission (Na).  The Ca {\sc ii} triplet in emission clearly
shows the double-peak structure expected from disc emission, and the
Ca {\sc ii} absorption features show the same phase delay as the Na
{\sc I} absorption lines.

The continua immediately adjacent to the lines of interest were fitted
(while masking the line) and then subtracted to give an average of
zero (as required by the Doppler tomography package, DOPPLER, written
by Tom Marsh). Spectra were then rebinned to a uniform velocity scale
with corresponding laboratory rest wavelengths as origins (the system
velocity is applied at a later stage). Spectra were then ready for use
in DOPPLER. It should be stressed that only the systemic velocity was
accounted for while making these Doppler maps; the 54.6 km s$^{-1}$
offset was not included.

\subsection{Doppler maps}

Figures 5--8 display the results of Doppler tomography on four of our regions of 
interest. Figure 5 shows the images created from H$\alpha$ emission. We can
clearly make out a circular disc, but most obvious is the intense
bright spot in the ($-V_{x}$, $+V_{y}$) quadrant. Since the region does 
not coincide with the secondary star Roche lobe, 
its origin must be due to a stream/disc interaction hot spot. 
The hot spot does not lie over either of the trajectories, but 
this is not unexpected. Collision of the
stream with the disc can produce a shock front as the stream's
ballistic motion is altered by mixing with the bulk motion of disc
material. Consequently, the velocity of matter that passes the shock front
should be a blend of the two, causing the emission region to lie
between the trajectories, just inside the outer edge of the disc, as observed
here. The same behaviour has also been seen in other systems (e.g., A0620--00;
Marsh, Robinson, \& Wood 1994; Neilsen, Steeghs, \& Vrtilek 2008). 
The spot also appears to be
extended along the trajectories,
which could be due to the matter stream penetrating deeply into the
disc before finally being deviated.

The predicted data
reproduce much of the original structure but appear to
eliminate the one-sided variation in both the double peak and $s$-wave
signals. This is an example of Doppler tomography not being able to
incorporate variable line flux into its modelling (this problem can be removed using modulation tomography; see Section 4.2). The variation we do
see in the predicted data arises from the asymmetric shape of the
hot spot, and rotational symmetry duplicates the variation in both
sides of the profile.

The disc itself appears to be circular, with no distinct features
other than the hot spot. Any asymmetric behaviour is better observed
in the right-hand column. Since it is inherently asymmetric, we
expected the hot spot to reappear in the asymmetric-component data and
maps, and it does so relatively unchanged. However, though the disc
appeared nearly uniform, we also continue to find some disc emission
in the asymmetric maps. While this is possibly a sign of real
structure, the minor levels of the remaining artifacts gives such a
theory little weight.

The trajectories provide us with a way to estimate both disc
dimensions and spot position; each marker represents a step of
$0.1R_{\rm L1}$ from the primary. For H$\alpha$ the ballistic
trajectory gives a position for the brightest spot region at $R =
(0.75 \pm 0.05)R_{\rm L1}$; however, if the spot follows the Keplerian
motion of the disc, the trajectory gives $R = (0.55 \pm 0.05)R_{\rm
  L1}$.  The inner edge of the hot spot should correspond to the outer
disc edge and appears to extend as far up the Keplerian trajectory as
$\sim0.8R_{\rm L1}$, close to the predicted outer disc truncation
radius ($\sim0.9R_{Roche}$ $\sim0.75R_{\rm L1}$).

\begin{table*}
\begin{footnotesize}
\vspace{-5pt}
\textbf{Doppler tomograms:} {Figures 5--8 are mosaics of six
  images. The left-hand column (from top to bottom) shows the trailed
  spectra used to construct the maps, the Doppler map, and the
  reconstructed data from the map model. The right-hand column shows
  the trail from which a simulated axisymmetric component has been
  removed, the resulting asymmetrical component map, and the
  reconstructed data. The simulated symmetric image is created by
  finding median values for annular portions of the original map,
  centered on the velocity of the primary (0, $-K_{1}$ km s$^{-1}$),
  and the predicted data/trail for this map is calculated and subtracted
  from the original data to give the asymmetric component. Overplotted
  onto the maps are the companion Roche lobe, predicted companion,
  primary and center-of-mass positions (crosses), and two lines
  representing the accretion stream ballistic trajectory (lower line)
  and the Keplerian velocity of the disc along the stream (upper
  line). The crosses along the trajectories represent steps of
  0.1R$_{\rm L1}$ from the primary and asterisks show the apsides of the
  stream.}
\vspace{-5pt}
\end{footnotesize}
\end{table*}

\begin{figure}
\centerline{\includegraphics[width=2.5in]{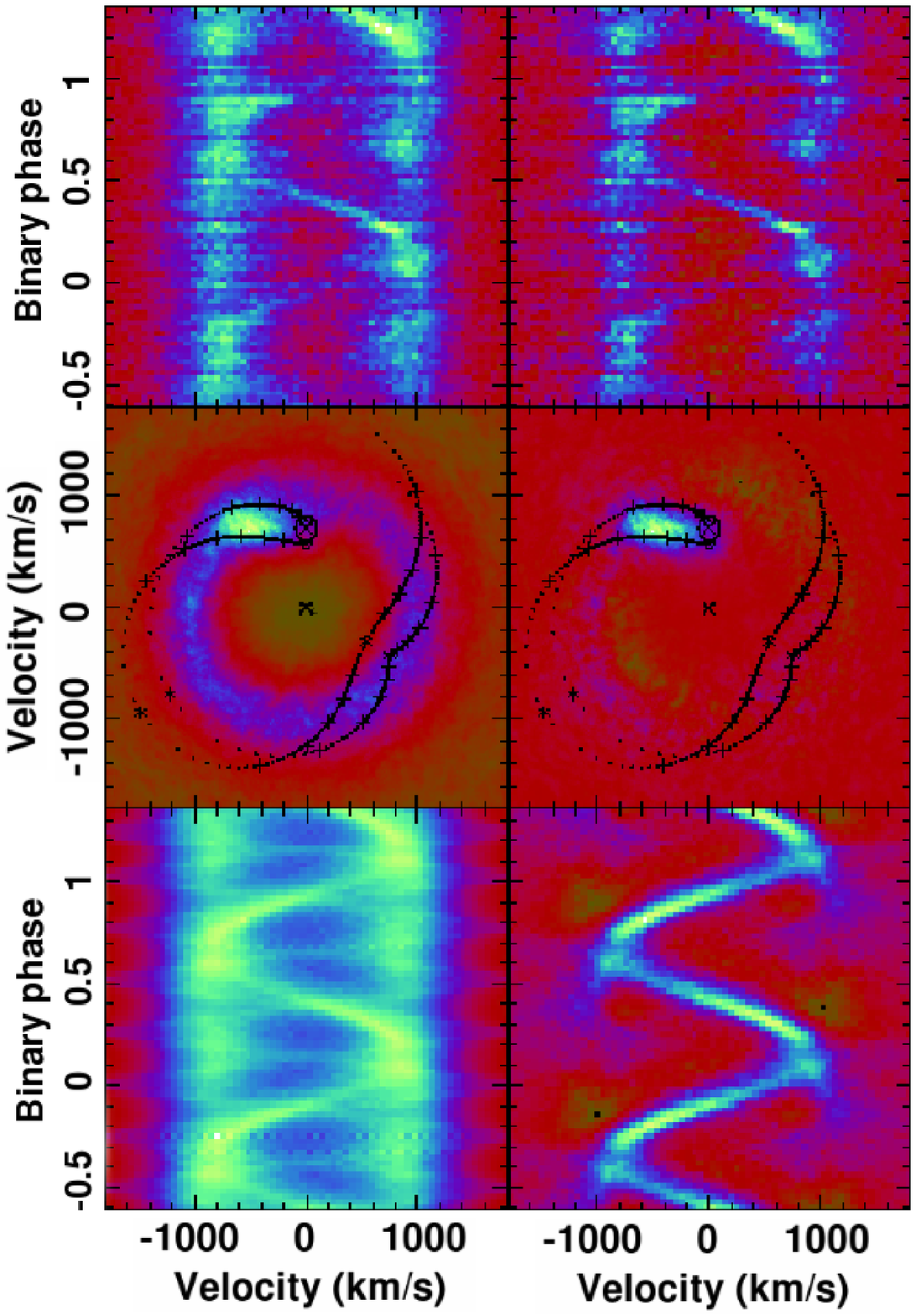}}
\vspace{-10pt}
\caption[H$\alpha$ Doppler tomograms.]{H$\alpha$ Doppler
tomograms. The hot spot can be seen between the two trajectories 
in both middle panels. The
circular disc is also clearly visible.}
\vspace{-0pt}
\centerline{\includegraphics[width=2.5in]{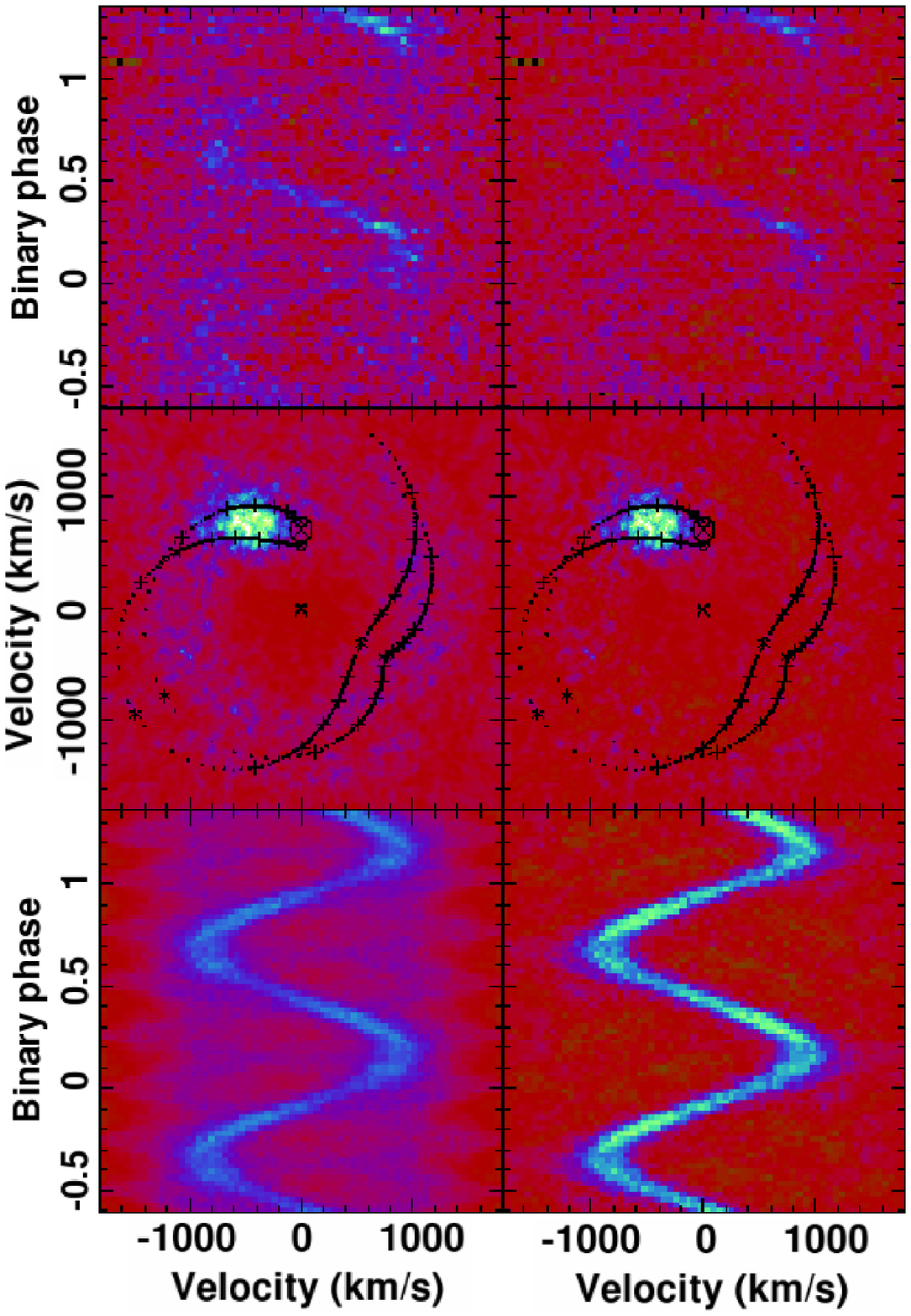}}
\vspace{-10pt}
\caption[H$\beta$ Doppler tomograms.]{H$\beta$ Doppler
tomograms. Again we observe the hot spot between the two trajectories,
but the disc is barely detected.}
\end{figure}

\begin{figure}
\centerline{\includegraphics[width=2.5in]{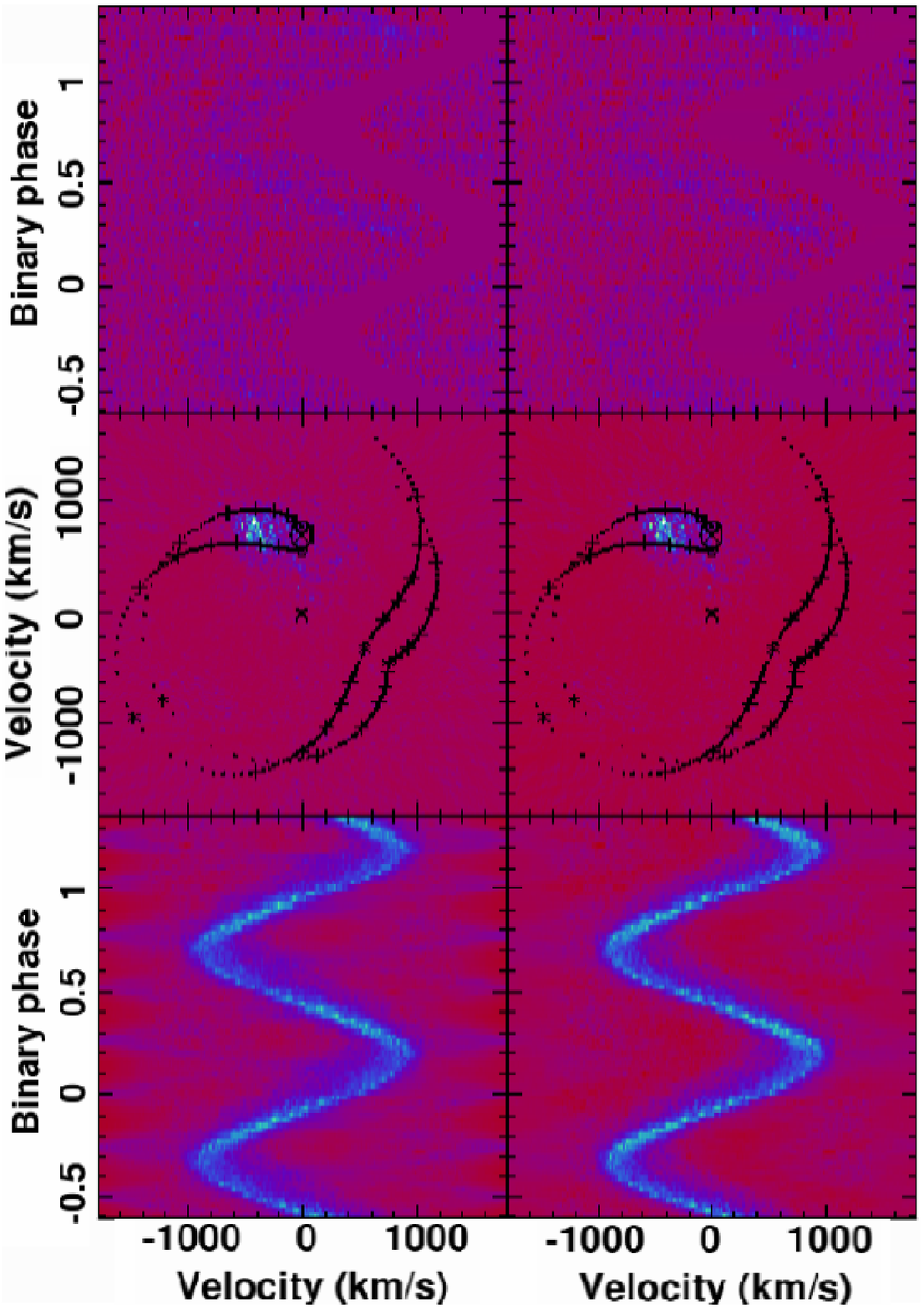}}
\vspace{-3pt}
\caption[He {\sc i} Doppler tomograms.]{He {\sc i} Doppler
tomograms. Only the hot spot is apparent.
\vspace{10pt}}
\centerline{\includegraphics[width=2.48in]{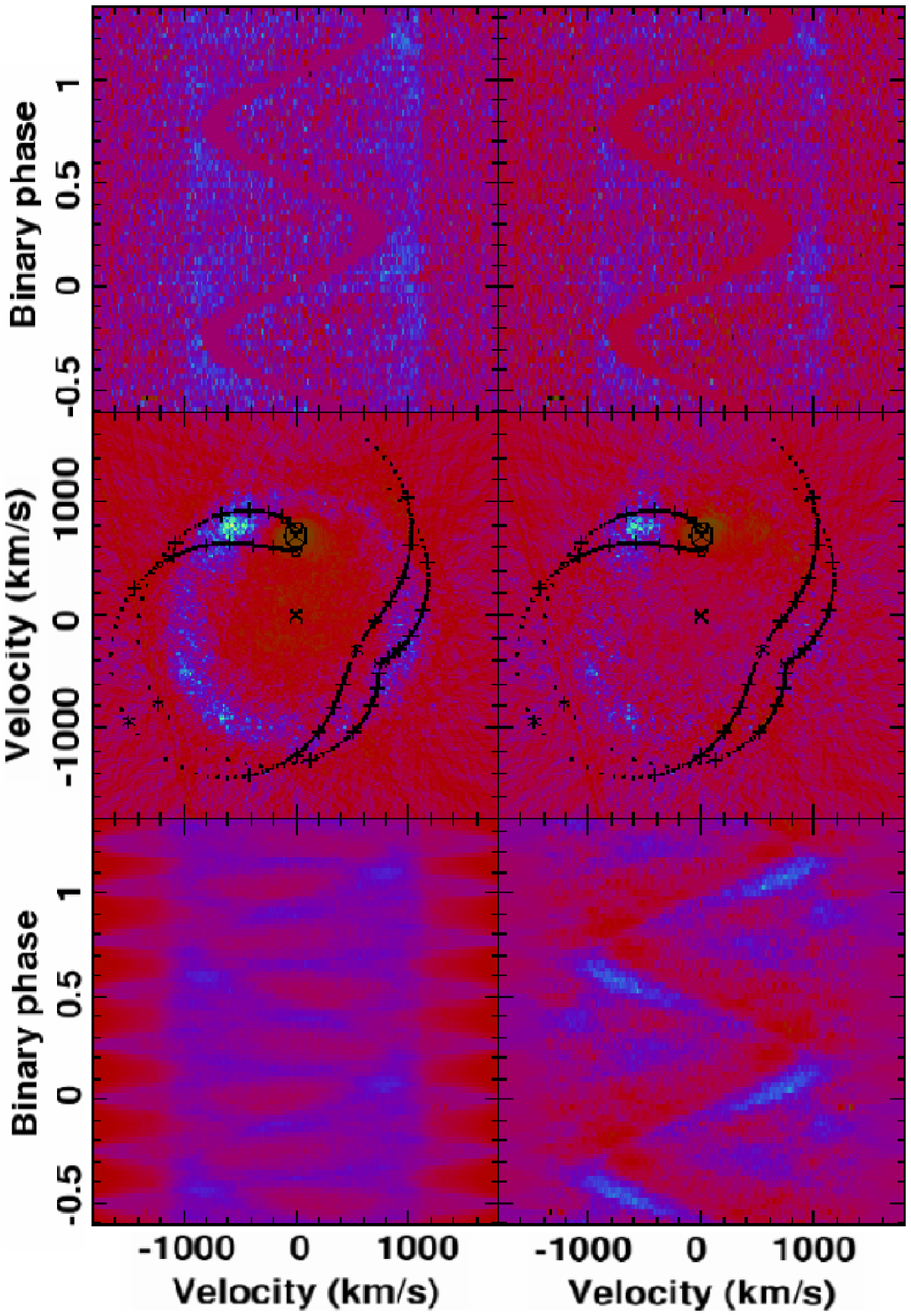}}
\vspace{-3pt}
\caption[Ca {\sc ii} $\lambda$8662 tomograms.]{Ca {\sc ii}
$\lambda$8662 tomograms. The disc and hot spot remain clear and
now we can also see absorption located at the secondary.}
\end{figure}

Figure 6 shows the tomograms and trails for H$\beta$. The weaker
double peak translates into minimal disc structure in the maps, but as
for H$\alpha$ what is visible of the disc remains relatively circular
and uniform. The H$\beta$ hot-spot position agrees very closely with
that of H$\alpha$, though appearing more circular. Figure 7 shows the
relevant images for He {\sc i}; being even weaker than H$\beta$, we can
only make out the $s$-wave in the data without any detectable 
double-peak structure and subsequently only the hotspot appears in the maps,
once again in the same position as the H lines. Since maximum-entropy
regularisation cannot deal with negative data points, in order to
accommodate the Na {\sc i} absorption line adjacent to He {\sc i} we
set the negative values to zero.  This does not significantly affect
the He {\sc i} image since the absorption does not overlap the
emission at any point.

In the case of Figure 8 (Ca {\sc ii}) we must make identical
alterations to allow for the Ca {\sc ii} absorption within the
emission pattern. While in this line there is some overlap between the
Ca {\sc ii} emission and absorption, it only augments the maps since
the absorption arises from the same transition as the emission,
allowing DOPPLER to accurately reproduce the velocity location of the
absorption.  The maps show both the disc and hot spot, as well as the
absorption region centered on the secondary's Roche lobe, confirming
the mass donor as the source of these absorption lines. The disc in
both the complete and asymmetric maps shows weak variation in
intensity at locations similar to those seen in H$\alpha$.

Finally, we have included tomograms of the four Na {\sc I} absorption
lines (Fig. 9). Spectra used for the images were created by zeroing
neighboring emission or absorption lines to leave a single absorption
$s$-wave (e.g., He {\sc i} and Na {\sc I} $ \lambda $5890 are reduced
to leave $\lambda $5896), and then the ordinate axis was inverted to
create a negative with which DOPPLER can work. The maps clearly show
the absorption centered on the centre of mass of the secondary. Hence,
the Na {\sc I} absorption seen in our spectra is a result of the
secondary, and shifts with time due to its orbital motion.

\begin{figure}
\centerline{\includegraphics[width=3.2in]{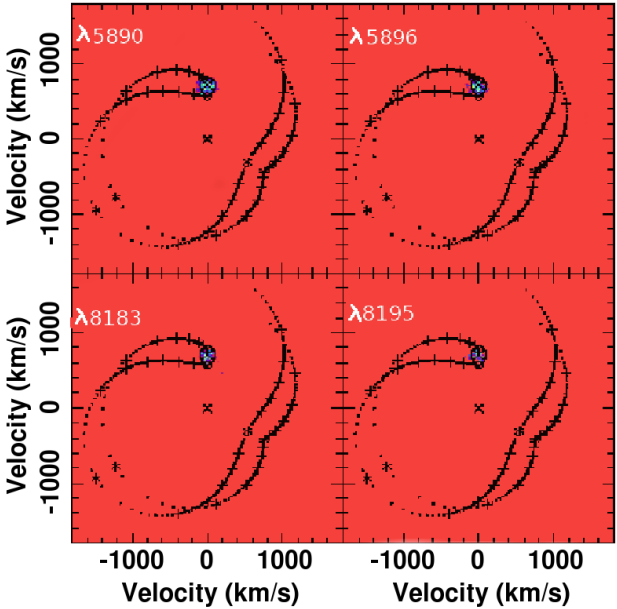}}
\caption[Na {\sc I} $\lambda$5890 absorption Doppler tomogram.]{Na {\sc I}
$\lambda$5890, $\lambda$5896, $\lambda$8183, and $\lambda$8195
absorption Doppler tomograms. The absorption is clearly
centered on the secondary's position in all the images. Overplotted features match those described for earlier tomograms (see the top of the previous page).}
\end{figure}

\subsection{Modulation tomography}

As we have seen, the intrinsic variations associated with the emission
of XTE J1118+480 can cause problems when producing Doppler
tomograms. Though the maps are clearly useful in identifying structure
within the disc, much of the behaviour we observe in the trails cannot
be represented accurately. Modulation tomography overcomes the need to
assume that the flux in the line remains constant over the orbit and
allows us to map fluctuations as a function of orbital period. Though
this makes the method more robust than Doppler tomography, it does
require data having higher signal-to-noise ratio; hence, we apply this
technique to the H$\alpha$ line alone. The process is described in
detail by Steeghs (2003).

We note that for tomography one needs an accurate measurement of the
systemic velocity of the binary system.  In our case the systemic
velocity is well constrained ($2.7 \pm 1$ km s$^{-1}$) using the many
absorption lines in the Keck data from the mass donor star (GH08).  In
addition, we tested the value of $\gamma$ determined by GH08 by
fitting the mean velocity for the sodium D absorption features and
find $2.1 \pm 2$ km s$^{-1}$.  The phase-invariant offset found in
H$\alpha$ line emission (discussed earlier) needs to be taken into
account when discussing disc structure since the visible disc's centre
may be offset compared to the primary via eccentricity. Modulation
tomography results in four tomograms rather than one: an average map
(akin to the original Doppler tomograms), a map detailing total
modulated emission and finally maps that show the individual sine and
cosine components of the modulated emission. The sine and cosine
labels correspond to how the emission varies in relation to the phase;
``sine'' modulated variations are those that form a sine profile
when plotted against phase (i.e., emission that peaks at phase 0.25 and
has a minimum at phase 0.75), while ``cosine'' modulated variation would
correspond to variations occurring in phase with a cosine profile
(i.e., maxima/minima occurring at phases of 0, 0.5, and 1). Variation with
a mixture of components are therefore those that appear in both maps.

We show modulation tomograms using the systemic velocity in Figure 10
and the offset velocity (in addition to systemic) in Figure 11. In
Figure 10, the average or ``constant'' map closely resembles the
original Doppler maps except the disc and hot spot are better defined,
and there appears to be a component of emission lying over the Roche
lobe of the secondary that was previously undetectable. The hot spot
remains elongated and in a similar position to that in the Doppler
maps, with an inner edge closer to $(0.75 \pm 0.05)R_{\rm L1}$ ($\sim$
0.9 $R_{\rm Roche}$); the estimated truncation radius (see Section
3.2). The color scale of the modulation maps represents a much smaller
dynamic range than that of the Doppler maps; the observed effects are
weak and do not necessarily dominate the behaviour of the overall
system.

\begin{figure*}
\centerline{\includegraphics[width=6.2in,angle=0]{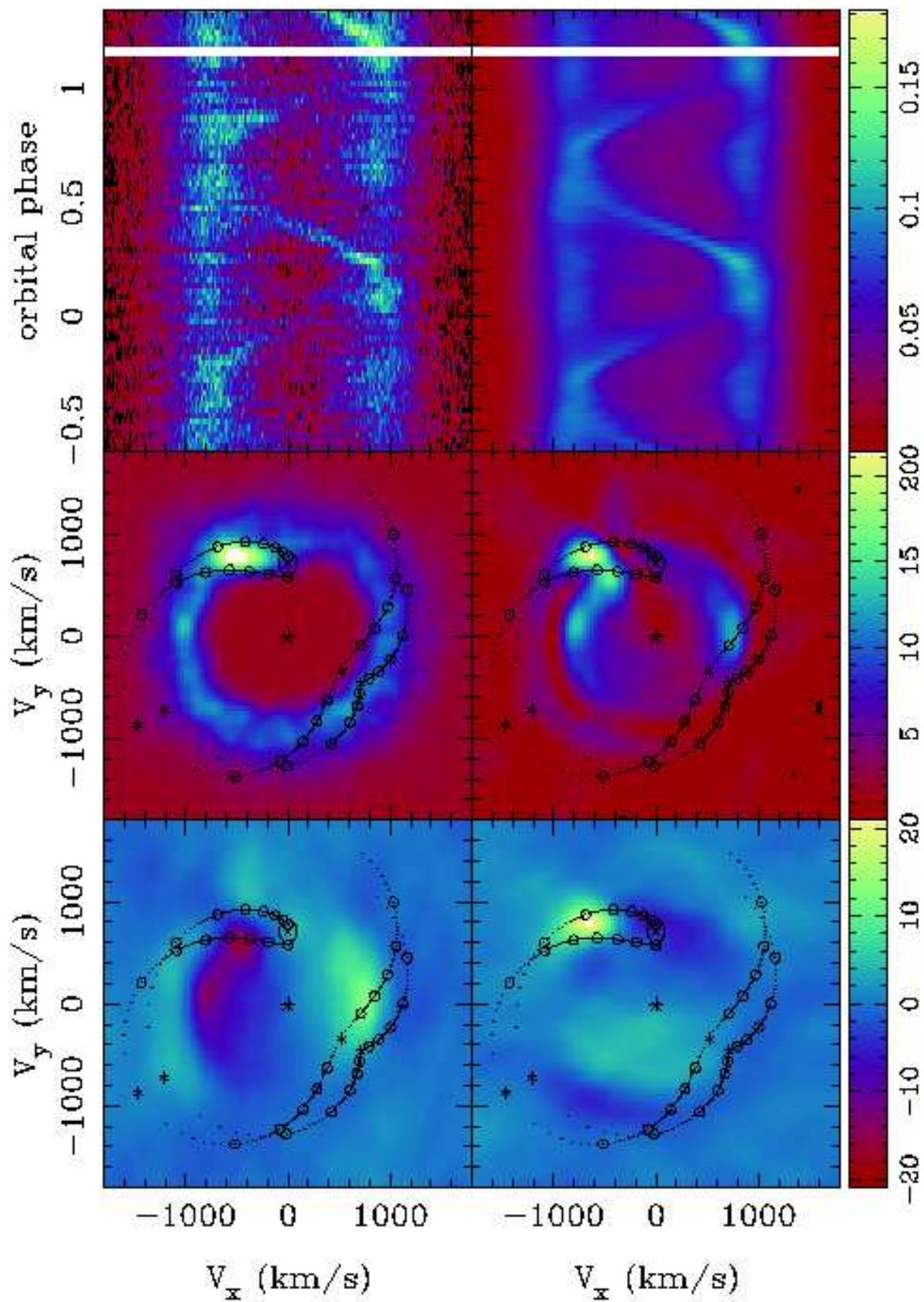}}
\caption[H$\alpha$ modulation tomograms.]{H$\alpha$ modulation
tomograms using $\gamma = 2.7$ km s$^{-1}$. The observed data 
(top left) are well reproduced by the
fitted data (top right).  The middle left-hand panel shows the constant
part of the disc with a strong hot spot, and the middle right-hand panel 
illustrates the amplitude of the
modulation. The bottom row represents the cosine (left) and
sine (right) components of rotation.}
\end{figure*}

\begin{figure*}
\centerline{\includegraphics[width=6.2in,angle=0]{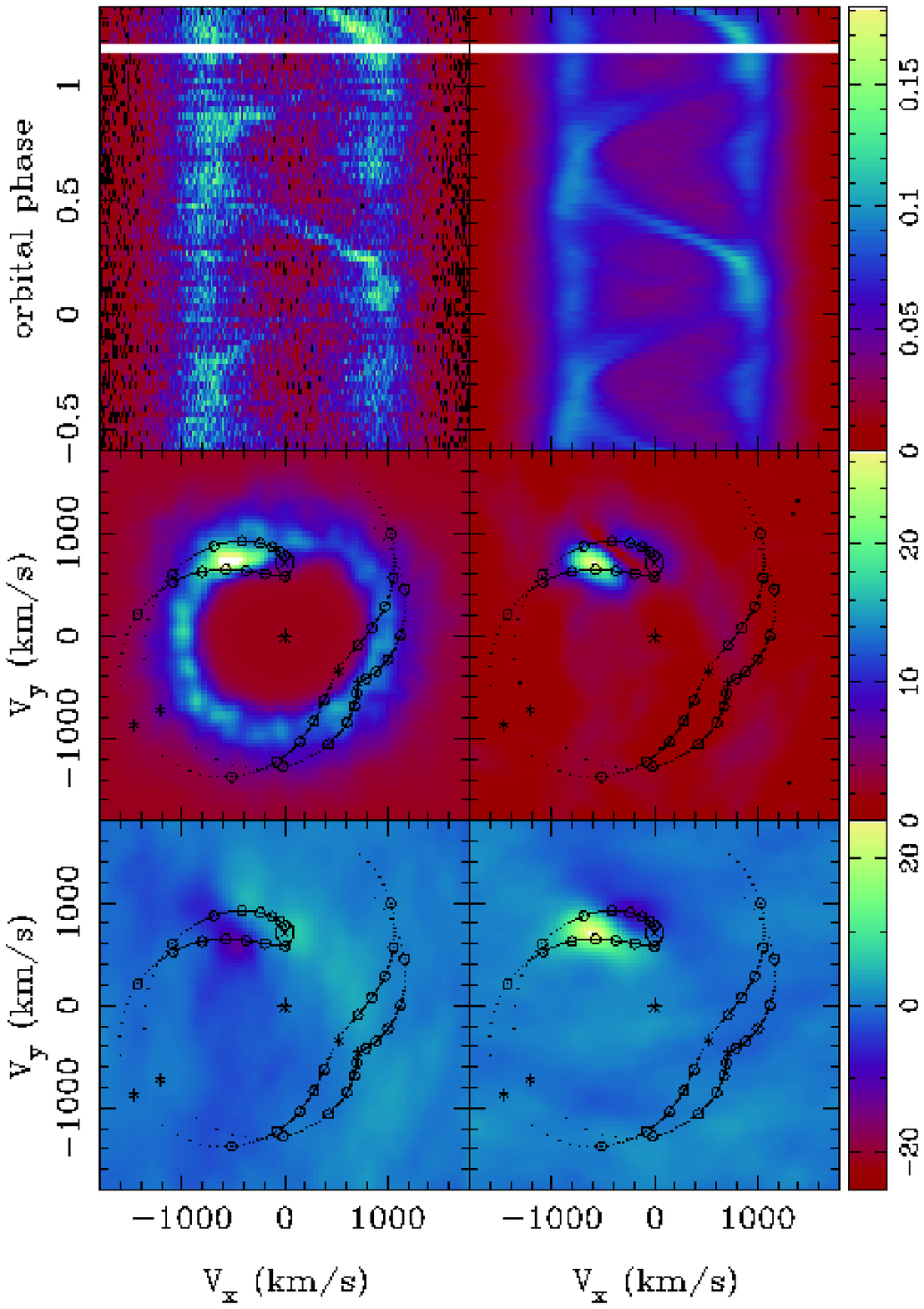}}
\caption[H$\alpha$ modulation tomograms.]{H$\alpha$ modulation
tomograms taking the measured centroid of the 
H$\alpha$ line as the systemic velocity ($\gamma = +57.3$ km s$^{-1}$). 
The stream for 
$q = 0.024$ is also plotted,
using $\gamma = 57.3$ km s$^{-1}$. The observed data 
(top left) are well reproduced by the
fitted data (top right).  The middle left-hand panel shows the constant
part of the disc with a strong hot spot, and the middle right-hand panel
illustrates the amplitude of the
modulation. The bottom row represents the cosine (left) and
sine (right) components of radiation.}
\end{figure*}

The hot spot appears in both cosine and sine component maps. Green and
red components of the spot in the cosine map represent regions which
brighten at opposite phases to each other, caused by exposure of
different faces of the extended stream-disc interaction as phase
varies. Both the green and red regions reside near $0.8R_{\rm L1}$
along the ballistic trajectory. The hot spot in the sine map is
dominated by a single green/white region that resides just above the
ballistic trajectory at $0.7R_{\rm L1}$.  The disc also appears to show
modulated variation in both component maps.  The cosine map shows
strong variation in the slower (inner map) disc regions that vary
opposite in phase. A similar but weaker crescent appears in the lower
half of the sine map. The crescent regions appear to be modulated at
the orbital period, as suggested by their positions and relative
amplitudes in both maps, rotated by 90 degrees in relation to each
other. At first this might imply a stationary (from an observer's
point of view) bright region which moves within the disc as the system
rotates, but we have yet to include the effects of the 54.6 km
s$^{-1}$ offset. The overall variation in the disc is easier to discern
in the final amplitude map (centre-right panel of Fig. 10). We can
see some disc variation at higher velocities in the lower-left
quadrant, but most is confined to the hot spot and slower disc
regions. However, due to the nature of the H$\alpha$ shift, these disc
structures are not real but once again a byproduct of not taking the
velocity offset into account.

Finally, notice that the modulated model data resemble the original
data far more closely than the unmodulated Doppler maps of Figure 5,
including the correct variation in $s$-wave and red-peak
intensities. The gap seen in the data is real and corresponds to the
larger than average gap in phase between observations 30 and 31 (see
Table 2). It is visible here (unlike in the Doppler tomography
mosaics) due to the higher resolution of the modulation tomography
plotting routines.

Figure 11 shows the modulation map which uses the measured centroid
offset of the H$\alpha$ line as the systemic velocity.  The accretion
disc appears remarkably symmetric, with little azimuthal structure in
the constant map and no significant modulation in the amplitude map.
The bright spot stands out nicely, and as can be seen from the plot,
varies in strength by about 25\%. The predominant positive sine term
reflects the fact that the hot spot is brightest near phase 0.25
(i.e., like the maximum of a sine wave).  The location of the hot spot
itself is stable, peaking at $R = 0.7 R_{\rm L1}$ on the ballistic
trajectory and with some mixing of disc velocities.  Though the spot
appears at different positions in the sine and cosine maps, this is not
a physical movement; the sine and cosine maps show how the anisotropic
emission from the stream-impact point causes the emission to modulate
over phase. Near phase 0.25 one looks more directly into the impact
site resulting in the peak in emission.  This disc appears relatively
stable and quiet (lacking any signs of eccentricity). Only a long
observing run covering the precession cycle could confirm if the disc
is able to remain circular and isotropic in its emission, yet still
precess.

\section{Discussion and Conclusions}

\subsection{XTE J1118+480 accretion in quiescence}

Tomograms constructed using H$\alpha$ as well as emission from
H$\beta$, He {\sc i} $\lambda$5876, and Ca {\sc ii} $\lambda$8662 each
reveal a well-defined hot spot lying, as expected, between the
ballistic stream and Keplerian disc velocity, where the accretion
stream hits the disc.  To04 saw no evidence for a hot spot in
H$\alpha$ during observations taken in 2003 January; instead, they
report that the asymmetric component of the map showed emission
centered on the secondary's Roche lobe, suggesting that mass transfer
had dropped significantly or ceased during the system's quiescent
period.  We propose that during our observation mass-transfer has
resumed (producing a hot spot) although there is still minimal
accretion onto the compact object, keeping the system in quiescence.

The peak-to-peak separation of the H$\alpha$ profile during our
observations is quite large and appears to be a sign of relative
quiescence.  During the rather large outburst in
Mar--Apr 2000 the peak-to-peak separation was at its lowest; during
the ``mini-outburst" of 2005 the peak-to-peak separation is somewhat
larger but still significantly smaller than during true quiescence.

\subsection{Imaging of the donor star and resolution of column 
density}

Because of its high Galactic latitude, the interstellar absorption
toward XTE J1118+480 is very low.  Garcia~\etal~(2000) limited
$E(B-V)$ to 0.024 mag. Using multiwavelength observations spanning the
energy range 0.4--160~keV, McClintock~\etal~(2001b) constrained the
column density to $N_{\rm H} = 1.2 \times 10^{20}$ cm$^{-2}$,
consistent with the Garcia~\etal~optical measurement.  Although
Dubus~\etal~(2001) claimed $N_{\rm H} = 1.8 \times 10^{20}$ cm$^{-2}$,
Chaty~\etal~(2003) showed that such high column densities are
inconsistent with accretion-disc models.  Dubus~\etal~(2001) measured
their column densities by using fits to Ca {\sc ii} absorption
features that they attributed to interstellar absorption based on
their spectral similarity to the Na {\sc I}~D lines.  Our tomograms
show unequivocally that the strong absorption features of Ca {\sc ii}
and Na {\sc I} are due entirely to the late-type secondary.  Doppler
maps of the absorption lines of Na {\sc I} $\lambda\lambda$ 5890,
5896, 8183, 8195 and Ca {\sc ii} $\lambda$8662 clearly demonstrate
their origin from the donor star.  Trailed spectra of many weaker
absorption features (some of which are visible in Fig. 4) show the
same $s$-wave pattern, indicating that they also originate from the
donor star.  We suggest that the high $N_{\rm H}$ measured by
Dubus~\etal~(2001) is incorrect because of their erroneous
interpretation of Ca {\sc ii} absorption features as being of
interstellar origin.

\subsection{Evidence for continuing precession and disc eccentricity}

Our Doppler tomograms show a disc that is relatively circular and
uniform in comparison to those presented by To02 and To04. For the
modulation maps which are more sensitive to effects that change over
the orbital period, asymmetric effects are only apparent when we use
the systemic velocity of the system.  The modulation effects
disappear when we use our measured velocity offset of 55 km s$^{-1}$
and instead we see a circular disc with isotropic emission.

Zu02 found a superhump period 0.3\% larger than the orbital period in
XTE J1118+480, implying a precession period of about 52~days.  Their
analysis of offsets in the centroid of H$\alpha$ emission were
consistent with a 52~day precession period.  H$\alpha$ offsets
measured by To04 up to two years after Zu02 showed that the precession
period persisted.  Gelino~\etal~(2006) and Shahbaz~\etal~(2005) did
not detect superhump periods, but their observations were not
sensitive to effects below the 0.5\% level.  Our measurement of an
offset in the H$\alpha$ centroid is consistent with effects from a
precessing eccentric disc with a precession period of $51.42 \pm
0.05$~days. However the fit requires the assumption that the precession
period remains stable over long periods of time and hence in various
system states, which is unprovable using existing offset measurements.

Zu02 compared XTE J1118+480 to SU UMa systems because of the superhump
effect they detected, but we note that in SU UMa systems superhump
effects are seen only in outburst.  In the case of XTE J1118+480, the
mass ratio ($q = 0.024$) is extreme enough to allow for effects of the
2:1 tidal resonance that is not possible with SU UMa dwarf novae
(Whitehurst \& King 1991).  Indeed, we find that the outer radius of
our disc extends well beyond the 3:1 and 2:1 tidal resonance radii,
supporting the continued presence of precession.

However, we are now left with the paradox of observing a circular,
uniform accretion disc that shows evidence of effects associated with
eccentricity, namely the emission-line offset. A physical scenario for
this would be a circular disc in which the matter's orbital velocity
varies with phase. The cause could be a slight offset in the disc
centre's position (i.e., not the same as the position of the primary's
centre of mass). This is obviously an unstable situation, and in turn
unlikely to be observed; however, we must consider two other factors
before discarding it as a solution. Firstly, compared to values
previously measured for the offset (e.g., 250 km s$^{-1}$ in Zu02) our
offset is small, which could in turn suggest a lower degree of
eccentricity required to cause it (i.e., too slight to detect). This is
difficult to confirm without additional observations covering more of
the precession cycle.  Secondly, it is possible we are simply not
observing enough of the disc at our chosen wavelengths to detect the
full extent of the eccentricity. Referring to the dissipation maps in
Figure 4 of Foulkes~\etal~(2004), were we ignorant of the inner half of
the disc, the remaining portion (especially the outer edge) would
appear very circular and yet have the required offset to fulfill the
above scenario. We may have essentially caught the system during a
short-lived arrangement.

\section{Summary}

We have presented emission-line analysis, Doppler tomography, and
modulation tomography of XRN XTE J1118+480 from Keck spectra taken on
the night of 2004 Feb. 14 when the system was in X-ray quiescence. All
emission tomograms reveal a well-defined hot spot lying between
ballistic stream and Keplerian disc velocity trajectories, and the
strongest emission lines clearly show the disc.  The hot spot implies
that mass transfer is ongoing, although accretion onto the compact
object is at a minimal level.  Applying the orbital solution from
GH08, tomography of several absorption lines allows us to reveal the
companion star.  The peak-to-peak velocity of the H$\alpha$ profile is
significantly larger during quiescence, as expected since the disc is
smaller.  Measurement of the H$\alpha$ offsets suggest the presence of
a precessing accretion disc.  Using our measured H$\alpha$ centroid
offset of 55 km s$^{-1}$ instead of the systemic velocity in
modulation mapping, anisotropic features are eliminated, producing a
relatively circular, isotropic disc.  Precession can be sustained
because the disc extends beyond the 3:1 and 2:1 tidal resonance radii,
but we need more observations at different precession phases in order
to determine how the offset we measure is affected by precession
effects and how they compare to those observed in the past.

\section{Acknowledgements}

The W. M. Keck Observatory is operated as a scientific partnership
among the California Institute of Technology, the University of
California, and NASA; it was made possible by the generous financial
support of the W. M. Keck Foundation. We thank the Keck staff, as well
as Ryan Chornock, for their assistance with the observations. 
The ASM data are provided by the ASM/RXTE teams at MIT and 
at the RXTE SOF and GOF at NASA's GSFC. We
acknowledge use of the MOLLY, DOPPLER, and TRAILER software developed
by Tom Marsh, MODMAP developed by Danny Steeghs, and the NIST Atomic
Spectra Database. This work was supported by NSF grants AST--0507637 to
SDV and AST--0607485
to AVF.  JIGH acknowledges support
from European Commission contract MEXT-CT-2004-014265 (CIFIST).

\end{document}